\documentclass[twocolumn,showpacs,preprintnumbers,amsmath,amssymb]{revtex4}

\usepackage{makeidx}
\usepackage{amsmath}
\usepackage{latexsym}
\usepackage{amssymb}
\usepackage{float}

\usepackage{graphicx}
\usepackage{dcolumn}
\usepackage{bm}

\begin{document}

\preprint{TOK-HEP061128}

\title{Angular Power Spectrum in Modular Invariant Inflation Model}

\author{Mitsuo J. Hayashi}
 \email{mhayashi@keyaki.cc.u-tokai.ac.jp}
\affiliation{
Department of Physics, Tokai University,\\
\mbox{1117~Kitakaname,~Hiratsuka, Kanagawa,~259-1292, Japan}
}

\author{Shiro Hirai}
\email{hirai@isc.osakac.ac.jp}
\affiliation{%
Department of Digital Games,
Osaka Electro-Communication University,\\
1130-70 Kiyotaki, Shijonawate, Osaka, 575-0063, Japan
}%
\author{Tomoyuki Takami}
\email{takami@isc.osakac.ac.jp}
\affiliation{%
Department of Digital Games,
Osaka Electro-Communication University,\\
1130-70 Kiyotaki, Shijonawate, Osaka, 575-0063, Japan
}%

\author{Yusuke Okame}%
 \email{5atrd001@keyaki.cc.u-tokai.ac.jp}
\affiliation{%
Graduate School of Science and Technology, Tokai University,\\
1117 Kitakaname, Hiratsuka, Kanagawa, 259-1292, Japan
}%
\author{Kenji Takagi}%
 \email{6atrd007@keyaki.cc.u-tokai.ac.jp}
\affiliation{%
Graduate School of Science and Technology, Tokai University,\\
1117 Kitakaname, Hiratsuka, Kanagawa, 259-1292, Japan
}%

\author{Tomoki Watanabe}%
 \email{tomoki@gravity.phys.waseda.ac.jp}
\affiliation{%
Advanced Research Institute for Science and Engineering, Waseda University,\\
3-4-1 Okubo, Shinjuku-ku, Tokyo, 169-8555, Japan
}%

\date{\today}

\begin{abstract}
We propose a scalar potential of inflation, motivated by modular invariant supergravity,
and compute the angular power spectra of the adiabatic density perturbations that result from this model.
The potential consists of three scalar fields, $S$, $Y$ and $T$, together with two free parameters.
By fitting the parameters to cosmological data at the fixed point $T=1$,
we find that the potential behaves like the single-field potential of $S$, which slowly rolls down
along the minimized trajectory in $Y$. We further show that the inflation predictions corresponding
to this potential provide a good fit to the recent three-year WMAP data, e.g. the spectral index $n_s = 0.951$. 


The $TT$ and $TE$ angular power spectra obtained from our model almost completely coincide
with the corresponding results obtained from the $\Lambda$CDM model.
We conclude that our model is considered to be an adequate theory of inflation that explains
the present data, although the theoretical basis of this model should be further explicated.  
\end{abstract}

\maketitle

\section{\label{sec:level1}Inflationary Cosmology}

Following three years of integration, the WMAP data has significantly improved\cite{ref:1}.
There have also been significant improvements in other astronomical data sets:
analysis of galaxy clustering in the SDSS\cite{ref:2,ref:3} and the completion of the
2dFGRS\cite{ref:2,ref:3};
improvements in small-scale CMB measurements\cite{ref:4}; much larger samples of high redshift
supernova\cite{highredshiftSN};
and significant improvements in lensing data\cite{lensing}.
The constraints on cosmological parameters, such as the spectral index and
its running, as well as the ratio of the tensor to the scalar, have also been improved.

The predictions of the theoretical inflation theories continue to find good agreement with these improved data sets. 
The $\Lambda$CDM model fits not only the three-year WMAP temperature and polarization data,
but also small scale CMB data, light element abundances, large scale structure observations,
and the supernova luminosity/distance relationship\cite{ref:1}.

As a favored scenario to explain the observational data,
it is customary to introduce a scalar field called inflaton into the theoretical models\cite{ref:6};
there are, however, several problems in constructing successful theories:
i) Precisely what is inflaton?
ii) What kind of theoretical frameworks are most appropriate to describe the theory of particle physics,
inflation and the recently observed accelerating universe?
iii) How can the contents of the universe be explained?: 
Baryonic matter 4\%, Dark matter 23\%, Dark energy 73\% and so on.
These problems seem to require a far richer structure of contents than that of the standard theory
of particles.
Furthermore, phenomenologically,
iv) Is the model consistent with the observed CMB angular power spectra?
In particular, inflaton should satisfy the slow-roll condition in order that the model
predicts the nearly scale-invariant spectral index, as well predicting a sufficiently large number of e-folds.
(See ref.\cite{ref:5} for a recent review of the theories of inflation.)

In this paper, we are proposing a potential which can give predictions consistent with those of the $\Lambda$CDM model.
This potential was originally derived by Ferrara {\it et al.}\cite{ref:11} in the context of $T$-duality and supersymmetry breaking
in string theory via gaugino condensation.
We have shown in ref.\cite{ref:9} that inflation and supersymmetry breaking can occur at the same time with an appropriate
choice of parameters.
However, because we would like to explore the phenomenological implications of this model as an inflationary theory,
we will suppose in the present study that the potential's parameters are not restricted by supergravitational backgrounds.

This paper is organized as follows:
In Sec. II, starting with a potential form which is modular invariant in $T$, we derive a stable inflationary trajectory
by fitting the parameters.
In Sec. III, we compute angular power spectra derived from this model, which are shown to agree
with three years of WMAP data, and almost coincide with
the corresponding results of the $\Lambda$CDM model.
In Sec. IV, we present the conclusions of this study, before offering a discussion of these results.
Finally in the Appendix, for completeness, we present an outline of the approach used to derive the present potential.
In this study we will use Planck units such that $m_{\rm pl}/\sqrt{8\pi}=1$.

\section{Inflationary Trajectory and Stability in variable $T$}

In order to construct the model, we introduce the inflaton field $S$ and the gauge-singlet complex
scalar fields $Y$ and $T$, motivated by the framework of modular invariant supergravity
conjectured from dimensionally reduced superstrings\cite{ref:7,ref:8,ref:9}.
However, we will not consider supergravitational backgrounds in this paper.

We consider a scalar potential for the three fields in the following form,
\begin{widetext}
\begin{eqnarray}
V(S,T,Y)
&=&
\frac{3(S+S^\ast)|Y|^4}{(T+T^\ast-|Y|^2)^2}
\Bigg(
	3b^2 \left| 1 + 3 \ln \left[ c \>e^{S/3b}\> Y \eta^2(T) \right] \right|^2
\nonumber\\
&&\hspace{-1.9cm}
	+ \frac{|Y|^2}{T+T^\ast-|Y|^2}
	\Bigg|
		S + S^\ast - 3b \ln \left[ c\>e^{S/3b}\>Y\eta^2(T) \right]
	\Bigg|^2
	+ 6b^2 |Y|^2
	\Bigg[
		2(T + T^\ast) \left| \frac{\eta^\prime (T)}{\eta(T)} \right|^2
		+ \frac{\eta^\prime (T)}{\eta(T)}
		+ \left( \frac{\eta^\prime (T)}{\eta(T)} \right)^\ast
	\Bigg]
\Bigg),
\label{V}
\end{eqnarray}
\end{widetext}
where $\eta$ is Dedekind's $\eta$-function, defined by
\begin{equation}
\eta(T)=e^{-2\pi T/24} \prod^{\infty}_{n=1}(1-e^{-2\pi nT}),
\end{equation}
$\eta '$ is its derivative with respect to $T$, 
and $c$ and $b$ are the free parameters of this potential.
For completeness, in the Appendix we will briefly review 
the outline of 
the derivation of Eq. (\ref{V}) in the context of supergravity
following ref.\cite{ref:11}.
In this context, $b$ is the one-loop renormalization group coefficient of gauge groups in
hidden sectors of ref.\cite{ref:11}, and is not free.
In the present study, however, in order to explain observed data using this potential form,
we treat $b$ along with $c$ as free parameters since we do not consider gaugino condensation.

The potential is modular invariant in the complex scalar field $T$, and is shown to be stationary
at the self-dual point $T=1$.
Such $T$-duality often plays important roles in various aspects of string theories\cite{polchinski};
e.g. this invariance is an unbroken symmetry at any order of string perturbation theory.
 One could therefore require the K\"aler potential and superpotential to be modular invariant.
For simplicity, $S$ and $Y$ are assumed to be real fields and the other matter fields are neglected.

One of the main purposes of this paper is to prove that the interrelation between $S$ and $Y$
gives rise to inflation.
As we will see later, upon minimizing $V$ with respect to $Y$ at the fixed point $T = 1$, one eventually finds
a single-field potential $V(S)$;  
the scalar field $S$ plays the role of the inflaton field.
Usually inflaton fields must satisfy the slow-roll condition in order for the corresponding inflation model to be successful.
Roughly speaking, $c$ determines the energy scale of $V$ while $b$ determines the flatness of $V$.
 If $c$ is small enough ($< 1$), the potential (\ref{V}) has no local minimum in $Y$ at fixed
$T$ other than the trivial minimum at $Y= 0$.
 We therefore have to choose the parameters $c$ and $b$ carefully.

We found that the potential $V(S,Y)$ at $T=1$ has a stable minimum at
$(Y_{\rm min},S_{\rm min})\sim (0.00877, 1.51)$ with $c=131,\ b=9.4$
(See Fig.1). These are the most suitable values for realizing the present experimental observations of three-year WMAP.
(If we could stick to $Y$ as a gaugino condensated scalar field, the local supersymmetry is broken at once with inflation,
providing a seed for observable supersymmetry breaking.
The value of $S_{\rm min}$ is consistent with $\langle S+S^\ast\rangle =\alpha^\prime m_{\rm pl}^2$.)
We can see inflation arises precisely due to the evolution of the scalar fields $S$ and $Y$ as follows:

\begin{figure}[h]
\begin{center}
\includegraphics[scale=0.8]{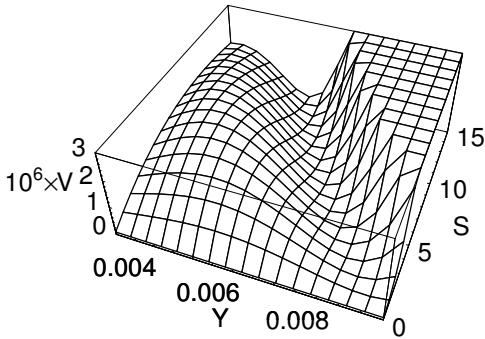}
\end{center}
\caption{
The potential $V(S,Y)$ at fixed $T=1$ (self-dual point) with $c=131,\ b=9.4$.
We can see a valley of the potential, $V_Y(S)=0$, and a stable minimum
at $(Y_{\rm min},S_{\rm min})\sim (0.00877, 1.51)$. 
}
\end{figure}

First, for the parameter values $c=131$ and $b=9.4$, the inflationary trajectory can
be well approximated by
\begin{equation}
Y_{\rm min}(S)\sim 0.00925 e^{-0.0355 S},
\end{equation}
which corresponds to the trajectory of the stable minimum for both $S$ and $Y$.

In Fig. 2, we have shown a plot of $V(S)$ minimized with respect to $Y$;
for large $c$, this potential is asymptotically approximated by 
\begin{equation}
V(S) \sim \frac{3}{2} \left( \frac{e^{-S/3b}}{a c} \right)^6
S (2 S^2 + 2 b S - b^2),
\end{equation}
where $a$ is a constant determined by $T$: $a \sim 0.824$ for $T = 1$.
Thus we have arrived at a single-field potential, starting from the modular invariant potential for the three fields.
The stability of the fixed point $T = 1$ will be shown later.


\begin{figure}[H]
\begin{center}
\includegraphics[scale=0.8]{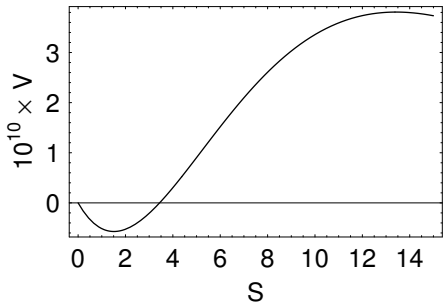}
\end{center}
\caption{
The potential $V(S)$ minimized with respect to $Y$.
The minimum value of the potential is $V(S_{\rm min})\sim -5.71\times 10^{-11}$.
}
\end{figure}

Next, the slow-roll parameters are defined by
\begin{equation}
\epsilon_\alpha = \frac{1}{2} \left( \frac{\partial_\alpha V}{V} \right)^2
\quad ,\quad 
\eta_{\alpha\beta} = \frac{\partial_\alpha \partial_\beta V}{V}.
\end{equation}
The slow-roll condition requires both these values to be less than 1.
 The end of the inflationary period is demarked by the slow-roll parameter $\epsilon_\alpha$ approaching the value 1.
Beyond the end of the inflationary period, ``matter" may be produced during the oscillations
around the minimum of the potential (reheating) at the critical density, i.e. $\Omega=1$.
Although any successful theory of inflation should explain the mechanism of the reheating process,
we postpone consideration of this reheating problem for later work.

For the present potential the values of $\epsilon_S$ and $\eta_{SS}$ are numerically obtained by
fixing the parameters $c=131\ {\rm and}\ b=9.4$ as shown in Fig. 3;
with these parameters the slow-roll condition is satisfied.

\begin{figure}[H]
\begin{center}
\includegraphics[scale=0.8]{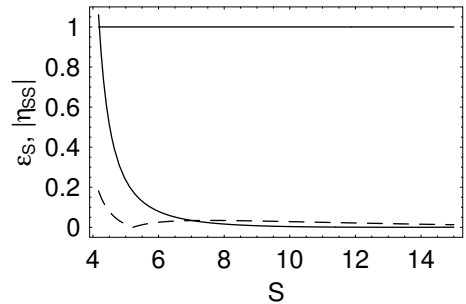}
\caption{
The evolution of the slow-roll parameters.
The solid curve represents $\epsilon_S$  while the dashed curve denotes $|\eta_{SS}|$. The plots
demonstrate that the potential $V(S)$ is sufficiently flat.
Inflation ends at $S \sim 4.19$ in our model.
}
\end{center}
\end{figure}

The potential $V$ is stable at the self-dual point $T=1$ in arbitrary points in the inflationary
trajectory for our choice of the parameters $c$ and $b$.
By choosing three points, i.e., the horizon exit, the end of inflation and the stable minimum,
and substituting the values of $S$, $Y$ at these points into the original $V(S,Y,T)$,
we will here demonstrate that the potential $V(T)$ has a minimum precisely at $T=1,$ and hence
is stable at these typical stages in the inflationary trajectory.
The variations of $V(T)$ are obtained numerically in Figs. 4 and 5 for the fixed parameters
$c=131$ and $b=9.4$.
 
\begin{figure}[H]
\begin{center}
\includegraphics[scale=0.8]{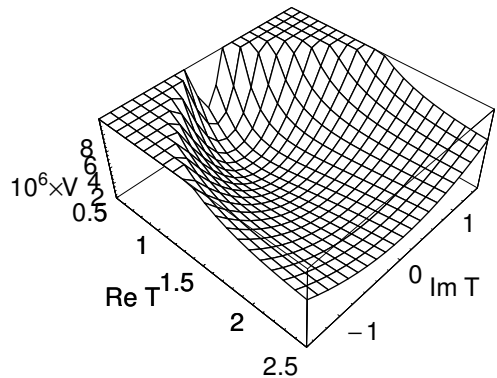}
\caption{
A 3D plot around the minimum of $V(T)$ as a function of the complex variable  $T$
for $S_{\rm min}$ and $Y_{\rm min}$.
Points along ${\rm Im}\ T = 0$ are obviously stable.
}
\end{center}
\end{figure}

\begin{figure}[H]
\begin{center}
\includegraphics[scale=0.8]{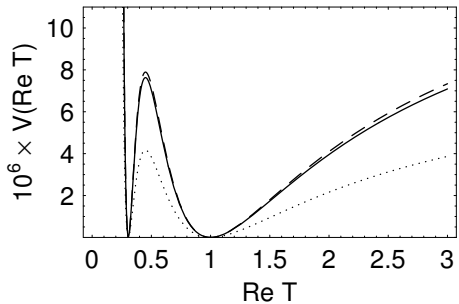}
\caption{
 Plots of $V(T)$ at ${\rm Im}\ T = 0$ at three representative inflationary stages.
It is easy to see that $T=1$ is stable.
The solid, dashed and dotted curves represent stages at the horizon exit, the end of inflation
and the stable minimum respectively.
}
\end{center}
\end{figure}

Now we have to verify the amount of inflation.
The number of $e$-folds at which a comoving scale $k$ crosses the Hubble scale $aH$ during
inflation is given by\cite{ref:6}
\begin{equation}
N(k) \sim
62 - \ln\frac{k}{a_0H_0} 
- \frac{1}{4} \ln \frac{(10^{16}\>{\rm GeV})^4}{V_k}
+ \frac{1}{4} \ln \frac{V_k}{V_{\rm end}},
\end{equation}
where we assume $V_{\rm end} = \rho_{\rm reh}$.
We focus on the scale $k_* = 0.05\> {\rm Mpc^{-1}}$ and the inflationary energy scale is
$V \sim 10^{-10} \sim (10^{16}\>{\rm GeV})^4$ as shown in Fig. 2.
 The number of $e$-folds which corresponds to our scale must therefore be around 57.

On the other hand, using the slow-roll approximation (SRA), $N$ is also given by
\begin{equation}
N \sim -\int^{S_2}_{S_1}\frac{V}{\partial V}dS.
\end{equation}
We could also have obtained the number of $e$-folds $\sim 57$, by fixing the parameters $c=131$ and $b=9.4$ and integrating from $S_{\rm end} \sim 4.19$
to $S_*\sim 11.6$,   i.e. our potential can produce
a cosmologically plausible number of $e$-folds.
Here $S_*$ is the value corresponding to $k_*$.

We can also compute the scalar spectral index and its running that describe the scale dependence
of the spectrum of the primordial density perturbation $\mathcal{P_R} = (H/\dot{S})^2 ( H/2\pi )^2$ \cite{ref:6, ref:19};
these indices are defined by
\begin{eqnarray}
n_s - 1 &=& \frac{d\ln \mathcal{P_R}}{d\ln k}, \\
\alpha_s &=& \frac{dn_s}{d\ln k}.
\end{eqnarray}
These are approximated in the slow-roll paradigm as
\begin{eqnarray}
n_s (S) &\sim& 1 - 6 \epsilon_S + 2 \eta_{SS}, \\
\alpha_s(S) &\sim& 16 \epsilon_S \eta_{SS} - 24 \epsilon_S^2 - 2\xi^2_{(3)},
\end{eqnarray}
where $\xi_{(3)}$ is an extra slow-roll parameter that includes the trivial third derivative of
the potential.
Substituting $S_*$ into these equations, we have $n_{s* }\sim 0.951$ and
$\alpha_{s*} \sim -2.50 \times 10^{-4}$.

Because $n_s$ is not equal to 1 and $\alpha_{s}$ is almost negligible, our model suggests
a tilted power law spectrum.
The value of $n_{s*}$ is consistent with the recent observations;
the best fit of three-year WMAP data using the power law $\Lambda$CDM model is
$n_s \sim 0.951$\cite{ref:1}.

Finally, estimating the spectrum $\mathcal{P_R}$ in SRA,
\begin{equation}
\mathcal{P_R}\sim\frac{1}{12\pi^2}\frac{V^3}{\partial V^2},
\end{equation}
we find $\mathcal{P_R}_* \sim 2.36\times10^{-9}$.
This result is also in excellent agreement with the measurements derived from observations.
It may also be noted that the energy scale $V\sim10^{-10}$ is also within the constrained range
obtained by Liddle and Leach\cite{ref:17}.
\\

Gravitational waves are an inevitable consequence of all inflation models.
 The tensor perturbation (the gravitational wave) spectrum is given by \cite{ref:6}
\begin{equation}
\mathcal{P}_{\rm grav} = 8 \left( \frac{H}{2\pi} \right)^2 = \frac{2}{3\pi^2}V.
\end{equation}
In SRA, the spectral index of $\mathcal{P}_{\rm grav}$ is given by the slow-roll parameters
$\epsilon$ and $\eta$ as
\begin{equation}
n_{T} = -2\epsilon.
\end{equation}
The relative amplitude of the gravitational waves and the adiabatic density perturbations is
given by
\begin{equation}
r = \frac{\mathcal{P_{\rm grav}}}{\mathcal{P}_{\mathcal{R}}}.
\end{equation}
The ratio $r'$ of the tensor quadrupole $C_2^{\rm grav}$ to the scalar quadrupole $C_2^{\mathcal{R}}$ is defined
by (Peiris {\it et al.} in \cite{ref:1})
\begin{equation}
r' = \frac{C_2^{\rm grav}}{C_2^{\mathcal{R}}} = 0.8625 r = 13.8 \epsilon,
\end{equation}
for the standard cold dark matter model.
The gravitational wave spectrum does not evolve and remains frozen-in as a massless field, even
after the horizon-exit, independent of the scalar perturbations\cite{ref:18}.
In contrast to this fact, the evolution of the primordial curvature fluctuation $\mathcal{R}$ is given by
the product of the transfer function $T_r(k)$ and $\mathcal{R}$:
\begin{equation}
\mathcal{R}_{\scriptsize\textbf{k}}^{(m)} = T_r(k) \mathcal{R}_{\scriptsize\textbf{k}}.
\end{equation}
Therefore, the ratio $r$ evolves as
\begin{equation}
\left(\frac{\mathcal{P}_{\rm grav}}{\mathcal{P}_{\mathcal{R}}}\right)^{(m)}
=-8T_r{}^2n_{T}
\end{equation}
up to the present time.
This result will be used in the calculation of the angular power spectra.

\section{The Angular Power Spectrum of the model}

Using our model, we can calculate the angular power spectrum to compare with WMAP analysis
and other experimental data\cite{ref:1,ref:2,ref:3,ref:4, highredshiftSN, lensing,hira}.
The multipoles $a_{lm}$ of the CMB anisotropy are defined by
\begin{eqnarray}
\Delta T &\equiv&
\frac{\delta T}{T}=\sum_{l>0}\sum_{m=-l}^{m=l}a_{lm}Y_{lm}(\bf{e}),
\\
a_{lm}&=&\int d\Omega_{\bf{n}} \Delta T( {\bf n}) Y_{lm}^*(\bf{e}),
\end{eqnarray}
where $Y_{lm}(\bf{n})$ are spherical harmonic functions evaluated in the direction $\bf{n}$.
The multipoles with $l\geq 2$ represent the intrinsic anisotropy of the CMB.
If the CMB temperature fluctuation $\Delta T$ is Gaussian distributed, then each $a_{lm}$ is an
independent Gaussian deviate with
\begin{equation}
\langle a_{lm} \rangle =0,
\end{equation}
and
\begin{equation}
\langle a_{lm}a_{l'm'}^* \rangle =\delta_{ll'}\delta_{mm'}C_l,
\end{equation}
where $C_l$ is the ensemble average power spectrum, or, the angular power spectrum of the CMB.
In general, the cosmological information is encoded in the standard deviations and correlations
of the coefficients:
\begin{equation}
\left \langle X Y \right \rangle = \left \langle a_{lm}^X a_{l'm'}^Y {}^* \right \rangle
= \delta_{ll'} \delta_{mm'} C_{l}^{XY}.
\end{equation}

For an arbitrary function $g(\bf{x})$, if we use a spherical expansion of the form
\begin{equation}
g({\bf x})=\int_0^\infty dk\sum_{lm}g_{lm}(k)\sqrt{\frac{2}{\pi}}kj_l(kx)Y_{lm}(\theta,\phi),
\end{equation}
where $j_l$ is the spherical Bessel function, and $(\theta, \phi)$ is the direction of ${\bf x}$,
then the angular power spectrum $C_l^{TT}$ and the temperature-polarization cross-power spectrum
$C_l^{TE}$ will be given by
\begin{eqnarray}
C_l^{TT} &=& 4\pi \int_0^\infty T_\Theta^2(k,l)\mathcal{P}_{\mathcal{R}}(k)\frac{dk}{k},
\\
C_l^{TE} &=& 4 \pi \int_0^\infty T_\Theta (k,l) T_E (k,l) \mathcal{P}_{\mathcal{R}} (k) \frac{dk}{k},
\end{eqnarray}
where $T_\Theta$ and $T_E$ are transfer functions and $\Theta$ is a brightness function.


Now we will describe the behavior of these power spectra according to our model.

The scalar spectral index is $n_s(k_*)=0.951$ and the running index is $\alpha_s(k_*)=-0.000250$
at $k_*=0.05 \ {\rm Mpc^{-1}}$, as has already been shown.
We consider a tensor-to-scalar ratio
$r' = 13.8 \epsilon =0.00910$ ($\epsilon = 0.000659$ at $k_* = 0.05 \ {\rm Mpc^{-1}}$).

We will use the CMBFAST\cite{cmbfast}, where we have assumed the cosmological parameters to be:
$\Omega_{\rm tot}=1$ for the total energy density,
$\omega =-1$ and $\Omega_\Lambda=0.762$ for the dark energy,
$\Omega_{\rm b}=0.0418$ and $\Omega_{\rm cdm}=0.1962$ for the baryonic and dark matter density,
$h=0.73$ for the Hubble constant.
The angular $TT$ power spectra were normalized with respect to 11 data points in the WMAP data
from $l=65$ to $l=210$, and the same values were used in the analysis of the angular $TE$ spectrum.

By using the likelihood method\cite{likelihood}, we calculated the $\chi^2$ values for the $TT$ and
$TE$ spectra, and also for their total sum. The results are shown in Table I.

The $\chi^2$ values for the $\Lambda$CDM model with $\tau =0.088$, which were given by
Spergel {\it et al}.\cite{ref:1}, were also calculated by the same method.
(For the one-year WMAP data, $\tau = 0.17$ was favored.)
On the other hand, the best fit of our model is realized at $\tau= 0.090$ for both $TT$ and $TE$
modes, which falls within experimental error,
while the $\chi^2$ values of our model seem to be better than those of the $\Lambda$CDM model.

The angular power spectrum of our model for the $TT$ mode at $\tau = 0.090$ are presented in Fig. 6.
Figs. 7 and 8 show the $TE$ spectra for $l\leq 50$ and with more detailed data $l \leq 20$
respectively also at $\tau = 0.090$.
Note that both the $TT$ and $TE$ spectra almost completely coincide with those of the $\Lambda$CDM
model across the whole range of $l$.
We would like to emphasize these two results as characteristic features of our model;
within the area of scalar inflation models, our model can be regarded as an alternative to the $\Lambda$CDM model. 

Although both our model and the $\Lambda$CDM model perform on the whole satisfactorily in explaining   the WMAP data,
there remains one inconsistency in the suppression of the spectrum at large angular scales
($l=2$)\cite{hira,ref:1}.
This problem is at present left to future investigations. 

In summary, the model we have here investigated is consistent with the present observational data
and the $\Lambda$CDM model.
\begin{table}[h]
	\begin{center}
		\caption{The $\chi^2$ values for the $TT$ and $TE$ spectrum and their total sum.
		The values of the $\Lambda$CDM model were calculated by using the results of Spergel {\it et al}.\cite{ref:1}.
		\vspace{0.5mm}}
		\begin{tabular}{c|c|c|c|c}
\hline
{} & $\tau$ & TT & TE & Total \\
\hline
\hline
Our model & {\ } 0.090 {\ }  & {\ } 1057.17 {\ } & {\ } 418.50 {\ } & {\ } 1475.67 {\ } \\
$\Lambda$CDM & 0.088 & 1057.56 & 418.50 & 1476.06 \\
\hline
		\end{tabular}
	\end{center}
\end{table}
\vspace{-0.7 cm}
\begin{figure}[H]
	\begin{center}
	\includegraphics[scale=0.45]{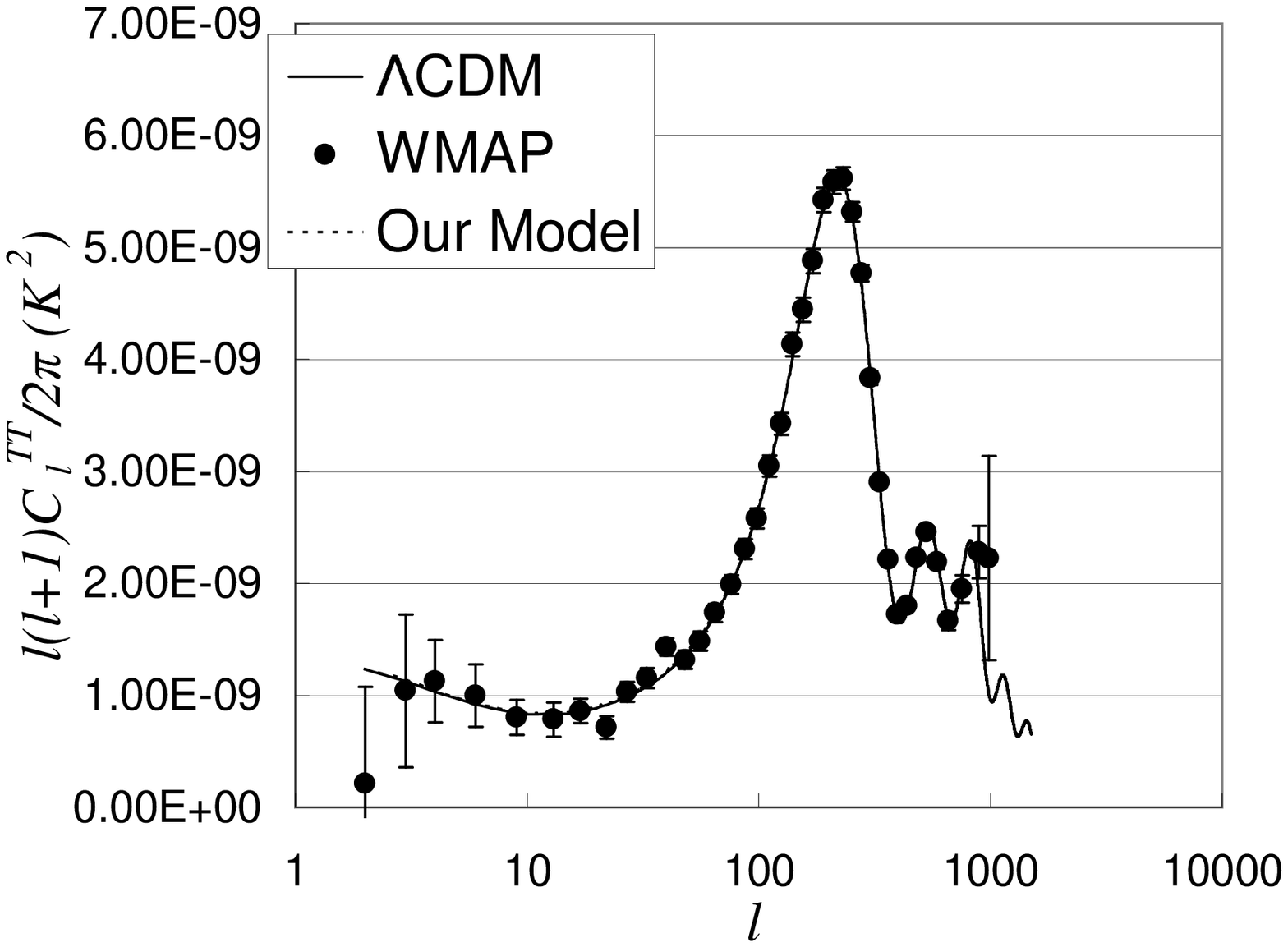}
	\caption{Temperature angular power spectrum ($TT$).}
	\end{center}
\end{figure}
\vspace{-1 cm}
\begin{figure}[h]
	\begin{center}
	\includegraphics[scale=0.42]{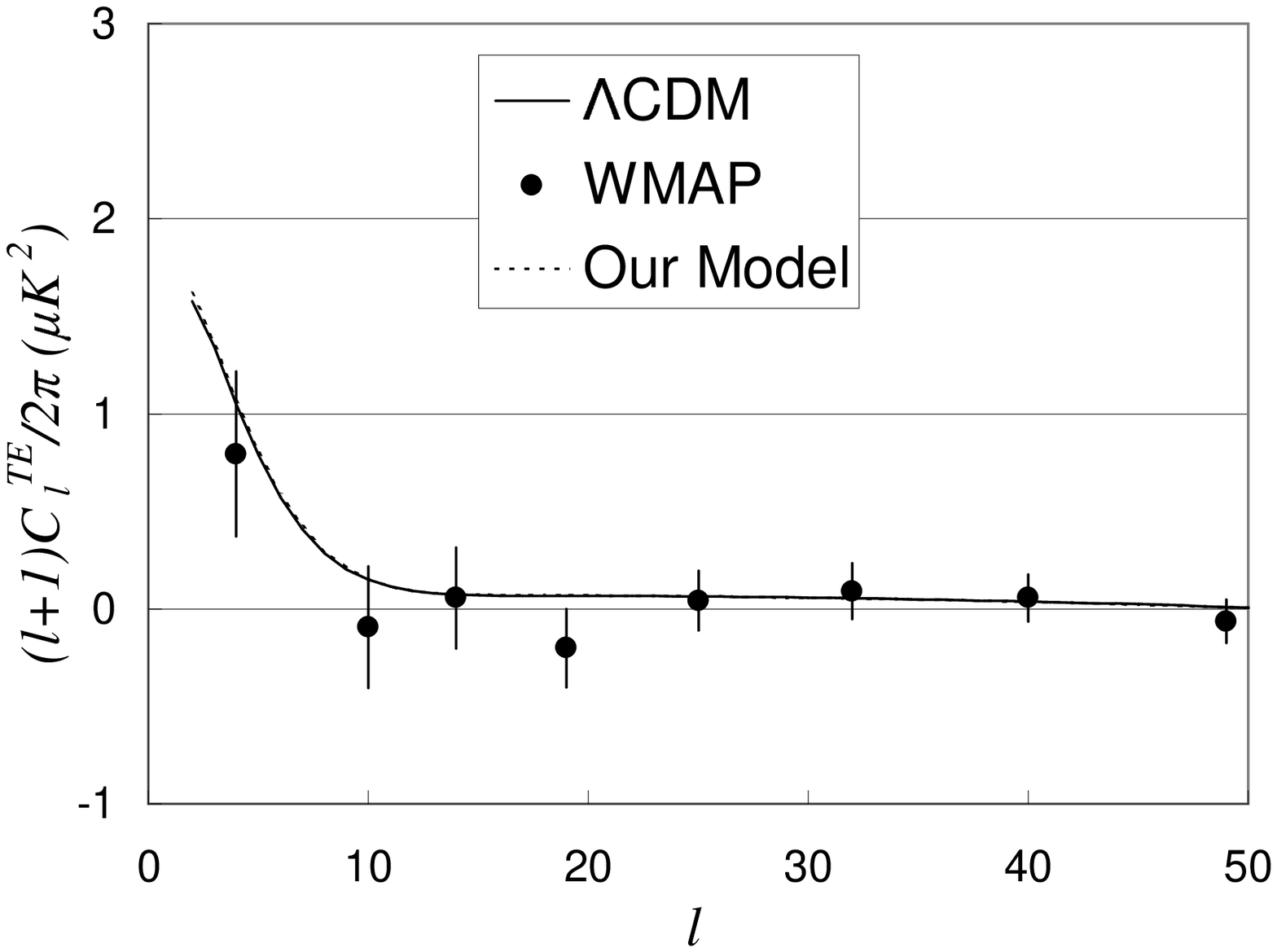}
	\caption{
	Temperature-polarization cross-power spectrum ($TE$) for $l < 50$.
	}
	\end{center}
\end{figure}
\begin{figure}[H]
	\begin{center}
	\includegraphics[scale=0.42]{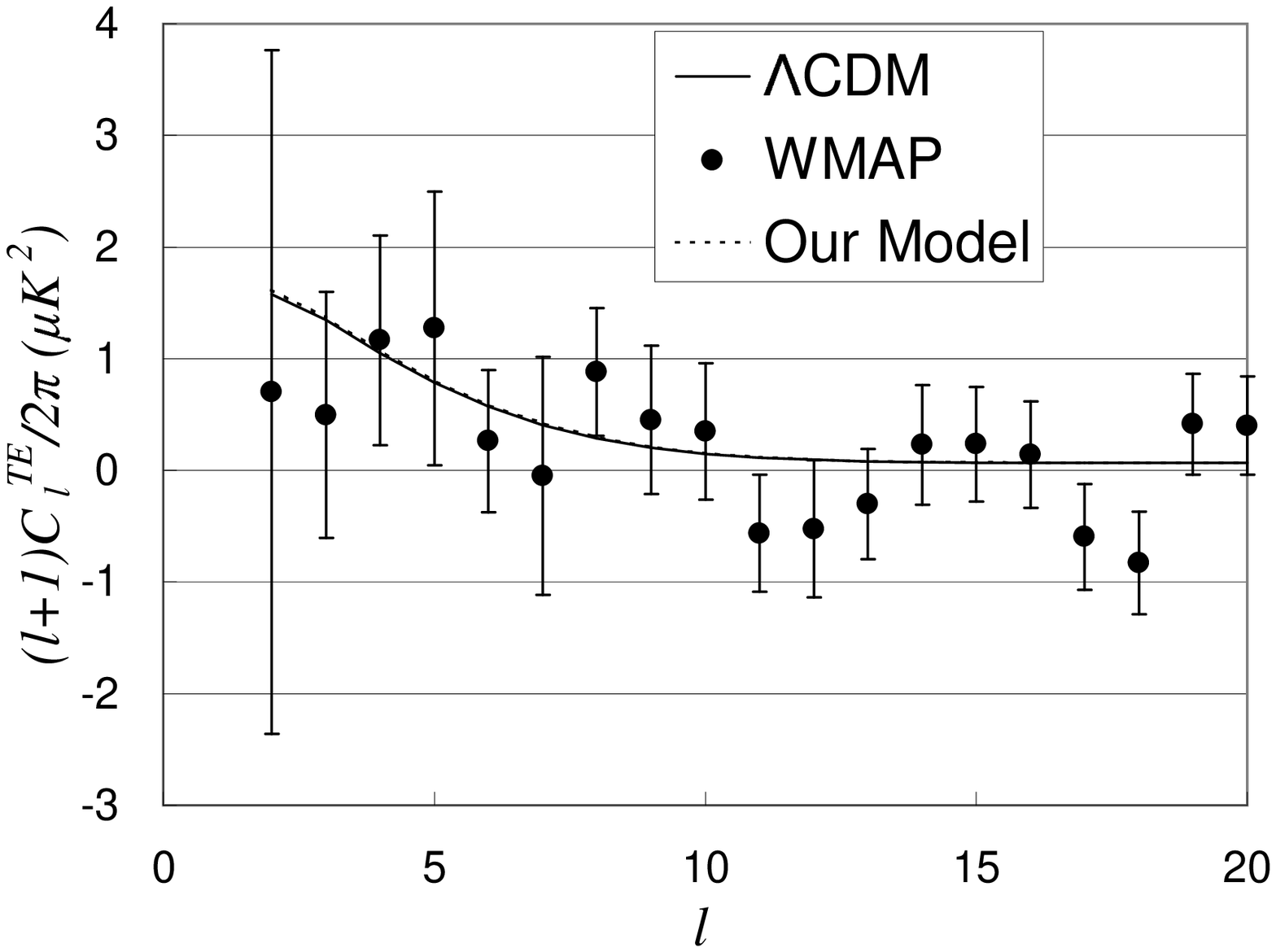}
	\caption{
	Temperature-polarization cross-power spectrum ($TE$) for $l < 20$ with more detailed data.
	For larger values of $l$ the results of our model are almost indistinguishable from those of the $\Lambda$CDM model.
	}
	\end{center}
\end{figure}

\section{Conclusion and Discussion}

We conclude that the results obtained with the present model are in good qualitative and
quantitative agreement with both the angular power spectra of the three-year WMAP
data, and the corresponding results obtained from the $\Lambda$CDM model.


Although our model started from considering a potential of three fields, it could eventually
be treated as a potential of a single field.
It is interesting to speculate whether the present model may be regarded as a model of multi-field
inflation, including the effects of fluctuations in the three fields.
From a particle physics point of view, it seems more natural to expect more than one scalar field
to roll during inflation.
In this case it may be necessary to consider a spectrum of isocurvature as well as curvature, and
the correlations between the two\cite{wands}.
Thus our model may be seen as a prelude to a full multi-field inflation model.

As we mentioned before, the inflaton potential we used was originally derived by Ferrara
{\it et al}. \cite{ref:11} in the context of dimensionally reduced supergravity from superstring
theories.
Their potential form was based on the idea of constructing an effective theory of gaugino
condensation\cite{ref:13}, incorporating the target-space duality,
where the gaugino condensation has been described by a duality-invariant effective action
for the gauge-singlet gaugino bound states $Y$ coupled to the fundamental fields as the dilaton $S$
and moduli $T$.
On the other hand, in ref.\cite{ref:10}, the gaugino-condensate has been replaced by its vacuum
expectation value to yield a duality-invariant ``truncated" action that depends on the fundamental
fields only.
The equivalence between these two approaches has been proved in refs.\cite{ref:12}.

It appears that supergravity is one of the most plausible frameworks to explain new physics,
including undetected objects, such as the inflaton, dark matter and dark energy.
In particular, since the inflaton field is concerned with Planck scale physics,
a dilaton field seems to be the most likely candidate for the inflaton.
The construction of a realistic supergravity is another problem that must be tackled in the future, and would appear to be a fruitful approach.

In the present study we have treated the parameters $b$ and $c$ as completely free parameters.
However, if we hope to consider our model as one given by string-inspired supergravity with
gaugino condensation of the hidden sector,
the value $b = 9.4$ is too large to be realistic, because $b$ is defined by $\frac{\beta}{96\pi^2}$
as the coefficient of the one-loop $\beta$-function of
the renormalization group equation for a gauge coupling constant of a gauge group, e.g. $E_8$.
Therefore our conclusion is restricted to presenting our model as a possible form of potential
which gives rise to an adequate inflation consistent with the present WMAP data.

Because the agreement with the WMAP observations does not seem merely accidental,
our next tasks include the investigation of an alternative derivation of an inflation potential of similar form to the present model using supergravitational theory.   Furthermore,
the reheating and matter production (dark and baryonic matters) following inflation will also be an immediate further area of investigation\cite{ref:5, matter prod}.

\appendix*
\section{derivation of the scalar potential}

In this appendix we briefly review the derivation of the scalar potential Eq. (\ref{V}) following Ferrara {\it et al.}\cite{ref:11}
in the context of modular invariant supergravity.
Assuming that the compactification of the superstring theory preserves $N=1$ supersymmetry, an effective theory should be of the general type of $N=1$ supergravity coupled to gauge and matter fields. 
The most general form of the Lagrangian in $N=1$ supergravity at the tree-level is\cite{ref:7}:
\begin{equation}
\mathcal{L}=
-\frac{1}{2}\left[e^{-K/3}S_0\bar{S}_0\right]_D
+\left[S_0^3W\right]_F
+\left[f_{ab}W^a_\alpha \epsilon^{\alpha\beta}W^b_\beta\right]_F,
\label{L}
\end{equation}
where the K\"{a}hler potential $K$ is given by
\begin{equation}
K=-\ln \left(S+S^\ast\right)-3\ln \left(T+T^\ast-|\Phi_i|^2\right),
\end{equation}
and the gauge function $f_{ab}$ is
\begin{equation}
f_{ab}=\delta_{ab} S.
\end{equation}
In order to construct an effective theory of gaugino condensation, we introduce the composite superfield $Y$
 of the gaugino condensation\cite{ref:11,ref:13}:
\begin{equation}
Y^3=
\delta_{ab}W^a_\alpha \epsilon^{\alpha\beta}W^b_\beta/S_0^3
=(\lambda\lambda+\cdots)/S_0^3,
\end{equation}
where $\lambda$ is the gaugino field in the Hidden sector.

The effective K\"{a}hler potential and superpotential incorporating modular invariant one-loop corrections are given by\cite{ref:11}
\begin{equation}
K=-\ln \!\left(S+S^\ast\right)
-3\ln \!\left(T+T^\ast-|Y|^2-|\Phi_i|^2\right),
\label{K}
\end{equation}
and
\begin{equation}
W=3bY^3\ln\left[c\>e^{S/3b}\>Y\eta^2(T)\right]+W_{\rm matter},
\label{W}
\end{equation}
where $\eta$ is Dedekind's $\eta$-function, $c$ is a free parameter in the theory and $b=\frac{\beta_0}{96\pi^2}$
($\beta_0$ is the one-loop beta-function coefficient).
Since $\langle S+S^\ast\rangle =\alpha^\prime m_{\rm pl}^2$, the choice
\begin{equation}
[e^{-K/3}S_0\bar{S}_0]_{\theta=\bar\theta=0}=[S+\bar{S}]_{\theta=\bar\theta=0},
\label{S}
\end{equation}
corresponds to the conventional normalization of the gravitational action:
\begin{equation}
\mathcal{L}_{\rm grav} \sim [e^{-K/3}S_0\bar{S}_0]_{\theta=\bar\theta=0}R.
\end{equation}

Then, after substituting Eqs.(\ref{K})-(\ref{S}) into Eq.(\ref{L}) and eliminating the auxiliary fields,
the scalar potential is obtained as Eq.(\ref{V}).

\end{document}